\documentclass[preprint,showpacs,showkeys,preprintnumbers,amssymb,nofootinbib
]{revtex4}
\usepackage{lscape}
\usepackage{graphicx}
\usepackage{dcolumn}
\usepackage{bm}
\usepackage{amsmath}
\usepackage[dvipsnames,usenames]{color}
\usepackage{mathtools}
\usepackage{enumitem,kantlipsum}
\usepackage{fancybox}
\def\beq{\begin{equation}}
\def\eeq{\end{equation}}
\def\beqna{\begin{eqnarray}}
\def\eeqna{\end{eqnarray}}
\def\eq(#1){(\ref{#1})}
\def\Eq(#1){Equation (\ref{#1})}
\def\eqs(#1)-(#2){(\ref{#1}) and (\ref{#2})}
\def\Eqs(#1)-(#2){Equations (\ref{#1}) and (\ref{#2})}
\def\eqss(#1)-(#2){(\ref{#1})-(\ref{#2})}
\def\Eqss(#1)-(#2){Equations (\ref{#1})-(\ref{#2})}
\def\fig#1{Fig.\ref{#1}}
\def\figs#1-#2{Figs.\ref{#1} and \ref{#2}}

\def\d#1{\!\!{\rm d}#1}
\def\dn#1#2{\!\!{\rm d}^{#1}{#2}\,}

\def\v#1{{\bf #1}}
\def\h#1{\hat{#1}}

\def\hv#1{\hat{{\bf #1}}}

\def\c#1{{\cal #1}}

\allowdisplaybreaks[2]
\begin{document}
\preprint{APS/123-QED}
\title{
Logical Fallacy of using the Electric Field\\ in Non-resonant Near-field Optics 
}
\author{Itsuki Banno$^{\ast}$}
\affiliation{
Graduate Faculty of Interdisciplinary Research Faculty of Engineering,
University of Yamanashi,\\
4-3-11 Takeda, Kofu, Yamanashi, 400-8511, Japan}
\email{banno@yamanashi.ac.jp}
\author{Motoichi Ohtsu}
\affiliation{
Research Origin for Dressed Photon, c/o Yokohama Technology Center, NICHIA Corporation, 3-13-19 Moriya-cho Kanagawa-ku, Yokohama-shi, 221-0022, Japan}
\date{\today}
\begin{abstract}
%
%
\pacs
{
78.67.-n, 
78.20.Bh, 
41.20.-q, 
42.25.Ja 
}
\keywords{
non-resonant condition, 
non-metallic material, 
optical near field, 
response function
}
We find that the electric field is not a suitable physical quantity to describe 
the response of {\it a non-metallic material} in the study of {\it non-resonant} near-field optics.
In practice, we show the spin-less one-electron two-level system  
responds differently to longitudinal and transverse electric fields under the non-resonant condition. 
This difference originates from {\it the non-relativistic nature} of the system,
and should exist in actual many-electron systems.
For this type of system, it is a logical fallacy to use the constitutive equation
in terms of the total electric field and the associated permittivity.
Recognizing this fallacy, 
both experimental and theoretical progress is needed in
the field of non-resonant near-field optics of non-metallic materials.
\end{abstract}
\maketitle
Under {\it non-resonant conditions} in the optical near field,
{\it non-metallic materials} cause various phenomena not 
observed in conventional optics, such as 
highly efficient light emission from 
indirect-transition-type semiconductors (LED\cite{SiLED,OhtsuSiLED} and 
Laser\cite{SiLaser,OhtsuSiLED}),
chemical reaction with insufficient photon energy 
(chemical vapor deposition\cite{NFCVD},  optical near-field lithography\cite{NFLitho}, 
optical near-field etching\cite{NFEtching}),
frequency up-conversion\cite{NFUpConv,PlsShpMs},
non-adiabatic effect beyond forbidden transition 
(local energy concentration\cite{nanoFountain},
nano-photonic gate device\cite{nanoPhotonicDevice}),
and gigantic magneto-optical rotation of the LED\cite{OhtsuSiLED,MO,MO2}.
Theoretically, {\it dressed photons}, namely, the localized electromagnetic field easily coupled with phonons, were introduced to allow non-adiabatic transitions\cite{DP01_Kawazoe,DP02_Kobayashi,OhtsuDP}.

This Rapid communication focuses on another fundamental role of the non-resonant condition in near-field optics 
(NFO) with non-metallic materials.
We examine the one-electron two-level system 
close to both the light source and the observation point under long wavelength approximation (LWA), 
and find it a logical fallacy to regard 
{\it the total electric field} as causing the response under the non-resonant condition.
In contrast, under the resonant condition or the far-field observation condition,
the electric field works as expected.
These findings originate from {\it the non-relativistic nature} of the system
and should be applicable in actual optical systems with non-metallic materials.
For the readability, calculation details are given in the last part of this paper.

\begin{figure}[tb]
\begin{center}
\includegraphics[width=0.5\textwidth]{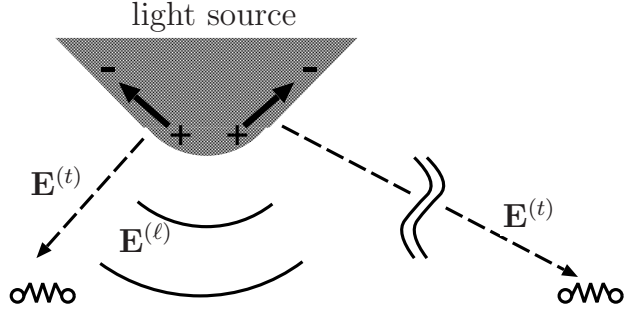}
\caption{
Target materials under near- and far-field incidences:
the former is exposed to the incident longitudinal and transverse electric fields 
simultaneously (the left side),
whereas the latter is exposed to only the transverse field (the right side).}
\label{fig:NFOvsOO}
\end{center}
\end{figure}
Suppose a small-scale material is placed in the vicinity of a nanostructure, which functions as a light source (\fig{fig:NFOvsOO}).
In such a system, under the NF incidence condition, the target material is exposed to 
longitudinal and transverse electric fields simultaneously, whereas in 
a system under the far-field incidence condition, the target material is 
exposed only to the transverse field, 
which survives far from the light source.
Therefore, the coexistence of longitudinal and transverse electric fields distinguishes 
such a system under the NF incidence condition from that under the far-field incidence condition.

Here, the longitudinal electric field originates from the charge 
density
on the nanostructure, obeys Coulomb's law, and  
has a non-radiative nature to localize around the nanostructure.
On the other hand, the transverse electric field originates from 
the transverse current density on the nanostructure, obeys 
the Ampere-Maxwell law and Faraday's law, and 
has a radiative nature allowing it to propagate far from the light source, accompanied by the magnetic field.
(The longitudinal current density is determined via the charge conservation law, once the charge density is known, and is not an independent source.)
Therefore, the two incidences coexisting in an NF optical system 
 have distinct properties. 

Furthermore, owing to {\it the non-relativistic nature} of the system,
the scalar and vector potentials appear in a different manner in the Hamiltonian, 
which governs the electron response, for example, \eq(H) of {\it Calculation details}\,\ref{cd:1} in the last part of this paper.
Considering that the scalar and vector potentials under the Coulomb gauge 
represent the longitudinal and transverse electric fields, respectively, 
one may confirm that 
the two types of incidences in NFO cause different responses.
Now our question is the following: under what condition can we observe these differences? 

\begin{table}[tbp]
\caption{
Classification of optical systems by distance 
from the target material to the light source
and distance from that to the observation point, together with 
a summary of the results; the validity of the electric field as the cause of the response. 
}
\label{table:NFOvsOO}
\begin{tabular}{| l | l | l |}\hline\hline
	                      & \bf Near-field observation   &\bf Far-field observation\\
	                      & \quad     Source:$\Delta\rho$ and $ \Delta \v{j}$ & \quad     Source: $\overline{\Delta \v{j}}$ \\	                     \hline\hline
$\displaystyle \begin{matrix*}[l]
\text{\bf Near-field incidence :}\\
\quad\Delta\v{E}^{(\ell)}+\Delta\v{E}^{(t)}
\end{matrix*}$
& \ovalbox{$\displaystyle \begin{matrix*}[l]
\text{\bf (I) NF optical system}\\
\text{non-resonant / resonant}
\end{matrix*}$}
&  \ovalbox{$\displaystyle \begin{matrix*}[l]
\text{\bf (I\hspace{-.1em}I) NF optical system}\\
\text{non-resonant / resonant}
\end{matrix*}$}
\\
\quad Validity of the electric field 
&\quad\quad\quad\quad \colorbox{black}{\textcolor{white}{NG}} / OK & \;\quad\quad\quad\quad OK / OK 
\\\hline
$\displaystyle \begin{matrix*}[l]
\text{\bf Far-field incidence :}\\
\quad \Delta\v{E}^{(t)}
\end{matrix*}$
& \ovalbox{$\displaystyle \begin{matrix*}[l]
\text{\bf (I$^{\prime}$) NF optical system}\\ 
\text{non-resonant / resonant}
\end{matrix*}$}
& \ovalbox{$\displaystyle \begin{matrix*}[l]
\text{\bf (I\hspace{-.1em}I$^{\prime}$\!){\footnotesize conventional optical\! system}}\!\\
\text{non-resonant / resonant}
\end{matrix*}$}
\\
\quad Validity of the  electric field
&\;\quad\quad\quad\quad OK / OK & \;\quad\quad\quad\quad OK / OK 
\\\hline\hline
\end{tabular}
\end{table}

Before proceeding with the analysis, let us first classify the optical systems. The two systems under near- and far-field incidence conditions 
in \fig{fig:NFOvsOO} are subdivided into two classes
depending on the near- or far-field observation condition.
These four classes are listed in Table \ref{table:NFOvsOO},  
together with a summary of the results mentioned below.
In particular, 
the systems of (I$^{\prime}$) and (I\hspace{-.1em}I$^{\prime}$)
are the limiting cases of null longitudinal incidence of 
the systems (I) and \quad(I\hspace{-.1em}I), respectively.
Thus, in the systems (I$^{\prime}$) and (I\hspace{-.1em}I$^{\prime}$),
the longitudinal response vanishes and 
the difference in response may not be observed.
In the following, therefore, we focus mainly on 
systems (I) and (I\hspace{-.1em}I),
in which longitudinal incidence exists.

\noindent
 {\bf Microscopic responses to longitudinal and transverse electric fields.}
%
Applying the linear response theory and the LWA
to the electron system of the target material on a small scale,  
the induced charge and current densities ({\it as a result of the response}), 
$\Delta\rho(\v{r},t)$ and $\Delta\v{j}(\v{r},t)$,
 are described as the total derivative with respect to
the longitudinal and transverse electric fields ({\it as the cause of the response}),
$\Delta \v{E}^{(\ell)}(\v{0},t)$ and $\Delta \v{E}^{(t)}(\v{0},t)$, 
where $\v{0}$ is the representative position in the electron system under the LWA:
\begin{align}
\label{rho_ind_av}
\Delta\rho (\v{r},t)
&= 
\chi^{\rho \leftarrow (\ell)}_{j}(\v{r},\omega) \,\Delta E^{(\ell)}_{j}(\v{0},t)\,
+
\chi^{\rho \leftarrow(t)}_{j}(\v{r},\omega) \,\Delta E^{(t)}_{j}(\v{0},t)\,,
\\
\label{j_ind_av}
\Delta j_{i} (\v{r},t)
&= 
\chi^{\v{j} \leftarrow (\ell)}_{ij}(\v{r},\omega)\,\Delta \dot{E}^{(\ell)}_{j}(\v{0},t)
+
\chi^{\v{j} \leftarrow (t)}_{ij}(\v{r},\omega)\,\Delta \dot{E}^{(t)}_{j}(\v{0},t)\,,
\end{align}
where the partial derivative coefficients, $\chi^{\cdots}_{\cdots}(\v{r},\omega)$'s
are susceptibilities (response functions), and
Einstein's rule is used for the summation over the vector indices,
for example, $\displaystyle \chi^{\rho \leftarrow (\ell)}_{j}(\v{r},\omega) \,\Delta E^{(\ell)}_{j}(\v{0},t)
 = \sum_{j=1}^{3} \chi^{\rho \leftarrow (\ell)}_{j}(\v{r},\omega) \,\Delta E^{(\ell)}_{j}(\v{0},t)\;$.
In \eq(j_ind_av), the time derivatives of the two types of electric fields, namely,
$\Delta \dot{E}^{(\ell)}_{j}(\v{0},t)$ and $\Delta \dot{E}^{(t)}_{j}(\v{0},t)$,
are regarded as the causes, instead of the two types of electric fields themselves.
The magnetic response vanishes in the leading order under the LWA
; see Ref.\cite{Cho02}.
The derivation of  \eqs(rho_ind_av)-(j_ind_av) is given in {\it Calculation details}\,\ref{cd:1}. 

For simple evaluation of the susceptibilities in \eqs(rho_ind_av)-(j_ind_av),
suppose we have a spinless one-electron system with two levels,
the ground and excited states in the non-perturbed system
with eigenenergies, $\hbar\omega_{0}$ and $\hbar\omega_{1}$, 
and orbitals, $\varphi_{0}(\v{r})$ and $\varphi_{1}(\v{r})$, respectively.
Those orbitals are assumed to be bound states expressed by real functions,
carry well-defined and distinct spatial parities (even and odd parities), 
and form the normalized orthogonal complete set.
The excitation energy is 
$\hbar\Delta\omega_{1}\equiv\hbar\omega_{1}-\hbar\omega_{0} \,>\,0$;
this finite excitation energy means that the target is a non-metallic material, such as
a molecule, nano-structured semiconductor and insulator. 

The susceptibilities in \eqs(rho_ind_av)-(j_ind_av) are derived in  {\it Calculation details}\,\ref{cd:2},
and those leading to the induced charge density result in the following:
\begin{align}
\label{suscept4rho01}
\chi^{\rho \leftarrow (\ell)}_{j}(\v{r},\omega) 
=&
\chi^{\rho \leftarrow (t)}_{j}(\v{r},\omega)
=
2q^{2} \, 
\frac{\eta}{\eta^{2}-1}\,
\frac{1}{\hbar\omega} 
\,\c{D}_{j}\,\varphi_{0}(\v{r}) \varphi_{1}(\v{r}) \,,
\\
\text{where} \quad\quad
\label{defeta}
\eta \equiv& \frac{\hbar \Delta\omega_{1}}{\hbar\omega} 
= \frac{\text{excitation energy}}{\text{photon energy}}\,,\;\text{and}
\\
\label{dtme}
\c{D}_{i}\equiv& \int \dn{3}{r} \: \varphi_{1}(\v{r}) \:  r_{i}\: \varphi_{0}(\v{r})\,.
\end{align}
This means that the responses to the longitudinal and transverse electric fields are common, 
such that the induced charge density has a linear relationship with
{\it the total electric field}, 
namely, 
$\displaystyle
\Delta\rho (\v{r},t) =
\chi^{\rho \leftarrow (\ell)\text{\:or\:}(t)}_{j}(\v{r},\omega) 
\left(
\Delta E^{(\ell)}_{j}(\v{0},t)+\Delta E^{(t)}_{j}(\v{0},t)
\right)
$.

The susceptibilities leading to the induced current density are not so simple and 
result in the following:
\begin{align}
\label{suscept4j0}
\chi^{\v{j} \leftarrow (\ell)}_{ij}(\v{r},\omega) 
=&
\frac{q^{2}\hbar^{2}\,}{m}\,
\frac{1}{\eta^{2}-1}\,
\frac{1} {(\hbar\omega)^{2}}\,
\c{D}_{j}\,
\left(    \partial_{i} \varphi_{1}(\v{r}) \varphi_{0}(\v{r}) - \varphi_{1}(\v{r}) \partial_{i} \varphi_{0}(\v{r}) 
\right)\,,
\\
\label{suscept4j1}
\chi^{\v{j} \leftarrow (t)}_{ij}(\v{r},\omega)
=&
\eta^{2}\,\chi^{\v{j} \leftarrow (\ell)}_{ij}(\v{r},\omega) 
- \frac{q^{2}\hbar^{2}\,}{m}\,
\frac{1} {(\hbar\omega)^{2}}\,
\varphi_{0}(\v{r})\varphi_{0}(\v{r})\,.
\end{align}
The susceptibility to the transverse electric field, \eq(suscept4j1), 
is composed of two terms.
The first term, namely, the resonant term, includes the energy denominator enhanced under the resonant condition, $\eta \simeq 1$, as in the susceptibility to the longitudinal electric field,
\eq(suscept4j0).
The second term, namely, the non-resonant term, does not include such a resonance factor.

%
\noindent
{\bf Equal responses under the resonant condition.}
%
Under the condition $\eta\simeq 1$ in all cases in Table \ref{table:NFOvsOO},
\eq(suscept4j1) is dominated by the resonant term (the first term) 
over the non-resonant term (the second term)
and asymptotically equals \eq(suscept4j0).
\begin{align}
\label{suscept4j01}
\chi^{\v{j} \leftarrow (t)}_{ij}(\v{r},\omega)\simeq \chi^{\v{j} \leftarrow (\ell)}_{ij}(\v{r},\omega)\,.
\end{align} 
\Eq(suscept4j01) together with \eq(suscept4rho01) reveal the equivalency of the responses 
to the longitudinal and transverse electric fields, so that 
the total electric field is regarded as the cause of the response in {\it any} optical system
under the resonant condition 
listed in Table \ref{table:NFOvsOO} .

%
\noindent
{\bf Equal responses under the far-field observation condition.}\;
%
In the system (I\hspace{-.1em}I) and (I\hspace{-.1em}I$^{\prime}$) in Table \ref{table:NFOvsOO},
the far field to be observed is insensitive to the details of the source but is determined by the
spatial average of the source.
Under the LWA, such an average can be achieved by the spatial average of the susceptibilities. 
Detailed calculations are shown in {\it Calculation details}\,\ref{cd:3}; the results are as follows.
\begin{align}
\label{suscept4rho_far}
\overline{\chi^{\rho \leftarrow (\ell)}_{j}(\v{r},\omega)} 
&=
\quad
\overline{\chi^{\rho \leftarrow (t)}_{j}(\v{r},\omega)}  =0\,,
\\
\label{suscept4j_far}
\overline{\chi^{\v{j} \leftarrow (\ell)}_{ij}(\v{r},\omega)}
&=
\overline{\chi^{\v{j} \leftarrow (t)}_{ij}(\v{r},\omega)}
=
\delta_{i\,j} \frac{q^{2}\hbar^{2}}{m\,\c{V}}  
\frac{1} {(\hbar\Delta \omega_{1})^{\,2}-(\hbar\omega)^{2}}\,,
\end{align}
where the overline represents the spatial average and $\c{V}$ is the
volume of the target material. 
From \eqs(suscept4rho_far)-(suscept4j_far), one may not observe 
different responses to the two types of incidences under the far-field observation condition.
The null response represented in \eq(suscept4rho_far) 
is reasonable because the induced charge density yields the longitudinal electric field,
which has a non-radiative nature and vanishes in the far-field regime.

\noindent
{\bf Unequal responses under the non-resonant, NF incidence, and NF observation conditions.}
%
The different responses 
claimed in the beginning of this Rapid communication
may be detected only in the system (I) in Table \ref{table:NFOvsOO}
under the non-resonant condition, which is just 
the compliment to the popular optical systems
under the resonant condition and/or the far-field observation condition.
In the NF optical system (I) with a non-metallic material under the non-resonant condition, 
{\it the total electric field}  is not the cause of the response;
therefore, the response may not be described by 
the macroscopic constitutive equation (MCE), namely,
the linear relationship between the polarization and "electric field" via permittivity,
and the microscopic susceptibilities are essential to treat separately 
the longitudinal and transverse incidences. 

In NFO, the response to the longitudinal electric field
is discussed in Chap. 5 in Ref.\cite{Cho02} and Chap. 9 in Ref.\cite{Keller01}.
The present work is a further comparison of the two responses, 
considering {\it the non-resonant condition}. 

The present model is very simple and the responses 
may be modified in a many-electron system or a low-symmetry system. 
However, 
the difference in the responses to the two types of electric fields 
originates in {\it the non-relativistic nature} 
of the system (as stated in the beginning of this Rapid communication),
and should survive in actual NF optical systems with non-metallic materials
(the materials with finite excitation energy). 
Actually, there is no reason for equating the two responses 
in the many-electron and low-symmetry systems.
Therefore, one may infer a guiding principle to highlight NF optical phenomena: 
under the non-resonant condition and simultaneous NF-incident and NF-observation conditions,  
non-metallic materials bring about NF-specific optical phenomena that may not be described 
by the MCE in terms of the electric field and the permittivity.
Some of the experiments mentioned in the beginning of this paper were 
performed under such conditions; thus,
we will analyze them in detail in future investigation.

\noindent
{\bf A remark on applying the finite differential time domain (FDTD) method to an NF optical system. }\;
The MCE in terms of the permittivity 
has been widely employed to calculate the optical near field 
in the FDTD method\cite{FDTD}.
One may notice that the permittivity in the FDTD method carries a simple spatial dependence 
and leads to some quantitative error.
Actually, the microscopic susceptibilities, for example, \eq(suscept4rho01), \eq(suscept4j0), and \eq(suscept4j1), have rippling spatial distributions 
originating from the orbitals.

In the case of the NF optical system (I) in Table \ref{table:NFOvsOO}
with a non-metallic material under the non-resonant condition,
the situation is more serious because 
the concept {\it electric field} is not available, such that
it is a logical fallacy to use the MCE. Thus, 
a novel simulation method is necessary.

\noindent
{\bf NFO and many-electron problem. }\;
Why has the comparison of responses to the two types of electric fields 
not been addressed in NF optical theory?
First, in the long history of optics, 
the NF optical system (I) in Table \ref{table:NFOvsOO} under a non-resonant condition 
has been out of focus. 
Such a system could not be resolved until 
the technical difficulty of NF observation was overcome. Additionally, resonance phenomena continue to
attract attention. Furthermore, 
even in NFO, there has been less emphasis on non-metallic materials, as opposed to metallic materials, which are essential for 
plasmonics.

The second reason is that the ordinary Hamiltonian for a many-electron system does not include the longitudinal electric field, which 
is rewritten to the two-body Coulomb interaction and eliminated.
With this Hamiltonian, the response to the longitudinal electric field incidence accompanies the Coulomb interaction, and is difficult to analyze.
Therefore, NFO is inevitably related to the many-electron problem; however, this has not been well recognized for a long time.
This study considered a one-electron system, avoiding the many-electron problem.
In future studies, 
the present scenario will be extended to a many-electron system and
nonlinear response, overcoming the many-electron problem,
and applying the findings to various phenomena mentioned in the
beginning of this Rapid communication. 

To the best of our knowledge, the present near-field optical system with non-metallic material
under the non-resonant condition is the third example that cannot be described in terms of electric field and/or magnetic field, after the superconductor system with the
Meissner effect and the electron system with the Aharonov-Bohm effect. 
The diversity of non-metallic materials including semiconductors, dielectrics, and magnetic materials
has been utilized in conventional optics. 
We believe that focusing on non-metallic materials in NFO promotes 
further development both conceptually and technically.

\noindent
{\bf Calculation details.}\;
%
Here we provide the calculation details, including the derivation of the unfamiliar relationship \eq(tradeoff) between 
two types of dipole transition matrix elements.

%
\begin{enumerate}[label= (\roman*), wide, labelwidth=!, labelindent=0pt]
\item
\label{cd:1}
\underline{Derivation of the microscopic constitutive equations, \eqs(rho_ind_av)-(j_ind_av).}
%
The incident scalar and vector potentials,
$\Delta\phi(\v{r},t)$ and $\Delta A_{i}(\v{r},t)$, are assumed to be 
monochromatic with the angular momentum $\omega$, 
and are expressed using the Coulomb gauge and LWA, as follows:
\begin{align}
\label{El}
\Delta\phi(\v{r},t)&=\Delta\phi(\v{r})\cos\omega t 
= \left( \Delta\phi(\v{0}) -  \Delta\v{E}^{(\ell)}(\v{0})\cdot\v{r} \right) \cos\omega t\,,
\\
\label{Et}
\Delta \v{A}(\v{r},t)&=\Delta\v{A}(\v{r})\sin(\omega t+\xi) = -\frac{1}{\omega} \Delta\v{E}^{(t)}(\v{0}) 
\sin(\omega t+\xi)\,,
\end{align}
where 
and $\xi$ is the phase difference between the two incident potentials.
The nanostructure is assumed to be a robust light source, which
is not affected by the target material, and the electromagnetic field is assumed to be a
classical field.

Using a spinless one-electron system, let us evaluate the induced charge and current densities caused by the coexisting incidences of the scalar and vector potentials.
The total Hamiltonian is as follows:
\begin{align}
\label{H}
& \h{H} =
\frac{1}{2m}
\left(
\frac{\hbar}{i} \frac{\partial}{\partial x_{i}(t)} -qA_{i}(\v{x}(t),t)\right)\left( \frac{\hbar}{i}\frac{\partial}{\partial x_{i}(t)} -qA_{i}(\v{x}(t),t) \right) 
+  q\phi(\v{x}(t),t)\,,
\end{align}
where $t$ is time, $\v{x}(t)$ is the position of the electron, 
and $q(=-e),m$ are the electron charge and mass, respectively.
The perturbation Hamiltonian is given by
\begin{align}
\label{ptbH}   
& \int \dn{3}{r} \left( \hat{\rho}(\v{r},t) \Delta\phi(\v{r},t) -  \h{j}_{i}(\v{r},t) \Delta A_{i}(\v{r},t)  \right),
\end{align}
where $\hat{\rho}(\v{r},t), \h{j}_{i}(\v{r},t) $ are the Heisenberg operators of
the charge and current densities defined as
\begin{align}
\label{rho}
\h{\rho}(\v{r},t) &=  q \delta^{3}(\v{r}-\v{x}(t))\, , 
\\
\label{j}
\h{j}_{i}(\v{r},t)&= 
\frac{q}{2m} 
 \left\{
 \left(\frac{\hbar}{i} \frac{\partial}{\partial x_{i}(t)}  -qA_{i}(\v{x}(t),t) \right)\delta^{3}(\v{r}-\v{x}(t)) 
 +  \delta^{3}(\v{r}-\v{x}(t))  \left(\frac{\hbar}{i} \frac{\partial}{\partial x_{i}(t)}  -qA_{i}(\v{x}(t),t) \right)
 \right\}\,.
\end{align}
The linear response theory leads to the operators of the induced charge and current densities, as follows: 
\begin{align}
\nonumber
\Delta\h{\rho}(\v{r},t)
=& \int_{-\infty}^{t} \!\!\!\d{t_{1}} \int \dn{3}{r_{1}} 
\left\{
\frac{1}{i\hbar} \left[ \h{\rho}^{(0)}(\v{r},t)\,,\, \h{\rho}^{(0)}(\v{r}_{1},t_{1}) \right]  \Delta\phi(\v{r}_{1},t_{1})
\right.
\\
\label{rho_ind}
&\left.
\hspace{0.15\textwidth}
-\frac{1}{i\hbar} \left[ \h{\rho}^{(0)}(\v{r},t)\,,\, \h{j}^{(0)}_{i_{1}}(\v{r}_{1},t_{1}) \right]  \Delta A_{i_{1}}(\v{r}_{1},t_{1})
\right\}\,,
\\
\nonumber
\Delta\h{j}_{i}(\v{r},t)
=&
\int_{-\infty}^{t} \!\!\!\d{t_{1}} \int \dn{3}{r_{1}} 
\left\{
\frac{1}{i\hbar} \left[ \h{j}^{(0)}_{i}(\v{r},t)\,,\, \h{\rho}^{(0)}(\v{r}_{1},t_{1}) \right]  \Delta\phi(\v{r}_{1},t_{1})
\right.
\\
\label{j_ind}
&\left.
\hspace{0.15\textwidth}
-\frac{1}{i\hbar} \left[ \h{j}^{(0)}_{i}(\v{r},t)\,,\, \h{j}^{(0)}_{i_{1}}(\v{r}_{1},t_{1}) \right]  \Delta A_{i_{1}}(\v{r}_{1},t_{1})
\right\}
 -\frac{q}{m} \h{\rho}^{(0)}(\v{r},t) \Delta A_{i}(\v{r},t) \,,
\end{align}
where $\h{\rho}^{(0)}$ and $ \hv{j}^{(0)}$ are the charge and 
current density operators, respectively, in the non-perturbed system.
The last term in \eq(j_ind) originates from {\it the non-relativistic nature} of the system
and is needed to maintain the charge conservation law.

Evaluating the expectation value using the ground state and
substituting \eqs(El)-(Et) leads to 
\eqs(rho_ind_av)-(j_ind_av), 
in which the causes of the responses are the two types of electric fields
and their temporal derivatives, defined as
\begin{align}
\label{delE}
&\Delta E^{(\ell)}_{j}(\v{0},t) \equiv \;\;\,\Delta E^{(\ell)}_{j}(\v{0}) \cos\omega t\,,\quad
\Delta E^{(t)}_{j}(\v{0},t) \equiv \;\;\,\Delta E^{(t)}_{j}(\v{0}) \cos(\omega t +\xi)\,,
\\
\label{delEd}
&\Delta \dot{E}^{(\ell)}_{j}(\v{0},t) \equiv \frac{\partial}{\partial t}\Delta E^{(\ell)}_{j}(\v{0},t)\,,
\quad\quad\;\,
\Delta \dot{E}^{(t)}_{j}(\v{0},t) \equiv 
\frac{\partial}{\partial t}\Delta E^{(t)}_{j}(\v{0},t)\,.
\end{align}

In the above, no magnetic response appears because it is
the higher order in the LWA\cite{Cho01,Cho02}. 
Cho derived a Taylor series of the non-local response function\cite{Cho03} under the LWA,
and assigned the electric permittivity and magnetic permeability in the MCE
as the term of order $\c{O}(ka)^{0}$ (the leading order) and $\c{O}(ka)^{2}$, respectively, 
where $ka \ll1$, $2\pi/k$ is the light wavelength, and $a$ is the representative size of the material. 

Furthermore, he pointed out that 
the MCE is irrational because the separability of the electric and magnetic responses 
and the term of order $\c{O}(ka)^{1}$ appears in a chiral symmetric system, including a NF optical system with a low-symmetric nanostructure.
The present work is concerned with another type of irrationality, which appears in the electric response (the leading order from the viewpoint of Cho) in NFO under {\it a non-resonant condition}.
\item
\label{cd:2}
\underline{Derivation of the expressions for susceptibilities,  \eq(suscept4rho01), 
\eqs(suscept4j0)-(suscept4j1).}\\
%
To obtain these formulas using the two-level model, 
we take the expectation values of \eqs(rho_ind)-(j_ind) using the ground state,
$\varphi_{0}(\v{r})$, and insert the projection operator [
the left side of the second equation in \eq(noc)],
assuming that the two orbitals are real functions, and form the normalized orthogonal complete set:
\begin{align}
\label{noc}
\int\dn{3}{r} \varphi_{m}(\v{r})\varphi_{n}(\v{r}) = \delta_{m\,n}\,,\quad\quad
\sum_{m} \varphi_{m}(\v{r})\varphi_{m}(\v{r}^{\prime}) = \delta^{3}(\v{r}-\v{r}^{\prime})\,,
\end{align} 
where $\varphi_{m}(\v{r})$ satisfies,  
\begin{align}
\label{states}
\h{H}^{(0)} \varphi_{m}(\v{r}) = \hbar\omega_{m}\,\varphi_{m}(\v{r})\,,\quad (m=0,1)\,.
\end{align}
Having real orbitals infers even temporal parity, such that
there is a null magnetic field in the non-perturbed system or
null vector potential in the non-perturbed Hamiltonian.
Furthermore, we use the well-known linear relationship between
the two types of dipole transition matrix elements,
\begin{align}
\label{wellknown}
 \c{C}_{i}&\equiv 
\int \dn{3}{r}\left( 
\partial_{i} \varphi_{1}(\v{r}) \varphi_{0}(\v{r}) - \varphi_{1}(\v{r}) \partial_{i} \varphi_{0}(\v{r}) 
\right)
= 
\frac{2m}{\hbar^{2}}\hbar\Delta\omega_{1}\, \c{D}_{i}\,.
\end{align}
\Eq(wellknown) is derived from the matrix element of the Heisenberg equation for dipole charge density:
\begin{align}
\frac{\partial}{\partial t}\, r_{j} \h{\rho}^{(0)}(\v{r},t)
= 
\frac{1}{i\hbar} 
\left[
 r_{j} \h{\rho}^{(0)}(\v{r},t)\,,\,\h{H}^{(0)}
\right] \,,
\end{align}
using $\displaystyle \h{\rho}^{(0)}(\v{r},t) = e^{-\frac{\h{H}^{(0)}t}{i\hbar}} \h{\rho}^{(0)}(\v{r},0)e^{+\frac{\h{H}^{(0)}t}{i\hbar}}$, the projection operator, \eqs(noc)-(states).
\item
\label{cd:3}
\underline{ 
Derivation of the spatial average of the susceptibilities, \eqs(suscept4rho_far)-(suscept4j_far).}
%
These following replacements in
\eq(suscept4rho01), \eqs(suscept4j0)-(suscept4j1)
lead to \eqs(suscept4rho_far)-(suscept4j_far):
\begin{align}
\varphi_{0}(\v{r})\varphi_{1}(\v{r})
&\quad\longrightarrow\quad 
\frac{1}{\c{V}}\int\dn{3}{r} \varphi_{0}(\v{r})\varphi_{1}(\v{r}) = 0\,,
\\
\partial_{i} \varphi_{1}(\v{r}) \varphi_{0}(\v{r}) - \varphi_{1}(\v{r}) \partial_{i} \varphi_{0}(\v{r})
&\quad\longrightarrow\quad
\frac{1}{\c{V}}\int\dn{3}{r} \partial_{i} \varphi_{1}(\v{r}) \varphi_{0}(\v{r}) - \varphi_{1}(\v{r}) \partial_{i} \varphi_{0}(\v{r})
=\frac{1}{\c{V}} \c{C}_{i}\,,
\\
\varphi_{0}(\v{r})\varphi_{0}(\v{r})
&\quad\longrightarrow\quad 
\frac{1}{\c{V}}\int\dn{3}{r} \varphi_{0}(\v{r})\varphi_{0}(\v{r}) =\frac{1}{\c{V}}\,.
\end{align}
To derive \eq(suscept4j_far), we additionally use the trade-off relationship
between the two types of dipole transition matrix elements,
\begin{align}
\label{tradeoff}
\c{D}_{i}\, \c{C}_{j} = \delta_{i\,j}.
\end{align}
This is effective in the two-level system with well-defined parity
and derived from the quantum-mechanical commutation relationship: 
\beq
\label{z-pz}
[ r_{i}\,,\, \frac{\hbar}{i}\partial_{j}  ] = i\hbar\,\delta_{ij}\,,\quad\text{i.e.,} \quad
r_{i} \left(   \frac{\hbar}{i}\partial_{j} \cdots \right) +  \frac{\hbar}{-i}\partial_{j} \left(r_{i} \cdots \right) = i\hbar \delta_{ij}\cdots \,. 
\eeq 
Inserting the projection operator 
between $r_{i}$ and $\frac{\hbar}{i}\partial_{j}$,
and eliminating the null integrals caused by mismatched parity result in \eq(tradeoff).
From \eq(wellknown) and \eq(tradeoff), $\c{D}_{i}$ and $\c{C}_{i}$ are specified as
\begin{align}
\label{CDspecify}
\c{D}_{i} = \frac{1}{\c{C}_{i}} = \frac{\hbar}{\sqrt{2m\,\hbar\Delta\omega_{1}}}\,.
\end{align}
(We do not use \eq(CDspecify) in this paper.)
\end{enumerate}
%
%
%
\begin{acknowledgements}
The authors are grateful Drs. I. Ojima (Research Origin of Dressed Photon (RODreP)), 
H. Sakuma (RODreP), H. Saigo (Nagahama Insitute of Bio-Science and Technology), 
K. Okamura (Nagoya Univ.),  H. Ando (Chiba Univ.) , and T. Kawazoe (Tokyo Denki Univ.) 
for useful discussions on the context of dressed photon.
One of the authors (I. B.) thanks Professor K. Cho in Osaka Univ.
for useful discussions concerning susceptibility.
This work is partially supported by JSPS KAKENHI Grant Number JP25610071 during 2013-2015, 
and Research Foundation for Opto-Science and Technology during 2018-2019.
\end{acknowledgements}
%

%
%

\begin{thebibliography}{00}
%
%
%
\bibitem{SiLED} T. Kawazoe, M. A. Mueed,  and M. Ohtsu, 
{Appl. Phys. B}  {\bf 104,} 747
(2011).
%
\bibitem{OhtsuSiLED} M. Ohtsu,
{\it Silicon Light-Emitting Diodes and Lasers} 
(Springer International Publishing, Switsland, 2016).
%
\bibitem{SiLaser} T. Kawazoe, M. Ohtsu, K. Akahane, and N. Yamamoto, 
{Appl. Phys. B}  {\bf 107,} 659
(2012).
%
\bibitem{NFCVD} T. Kawazoe, Y. Yamamoto, and M. Ohtsu, 
{Appl. Phys. Lett.} {\bf 79,} 1184
(2001).
%
\bibitem{NFLitho} H. Yonemitsu, T. Kawazoe, K. Kobayashi, and M. Ohtsu, 
{J. Photolumin.} {\bf 122,} 230
(2007).
%
%
\bibitem{NFEtching} T. Yatsui, K. Hirata, W. Nomura, Y. Tabata, and M. Ohtsu, 
{Appl. Phys. B}  {\bf 93,} 55
(2008).
%
\bibitem{NFUpConv} T. Kawazoe, H. Fujiwara, K. Kobayashi, and M. Ohtsu,  
{IEEE J. of Selected Topics in Quantum Electronics} {\bf 15,} 1380
(2009).
%
\bibitem{PlsShpMs}
H. Fujiwara, T. Kawazoe, and M. Ohtsu,
{Appl. Phys. B} {\bf 100,} 85
(2010).
%
\bibitem{nanoFountain}
T. Kawazoe, K. Kobayashi, and M. Ohtsu,
{Appl. Phys. Lett.} {\bf 86,} 103102-1
(2005).
%
\bibitem{nanoPhotonicDevice}
T. Kawazoe, M. Ohtsu, S. Aso, Y. Sawado, Y. Hosoda, K. Yoshizawa, K. Akahane, N. Yamamoto, and M. Naruse, 
{Appl. Phys. B} {\bf103,} 537
(2011). 
%
\bibitem{MO}
N. Tate, T. Kawazoe, W. Nomura, and M. Ohtsu,
{Scientific Reports} {\bf 5,} 12762-1
(2015) .
%
\bibitem{MO2}
N. Tate, T. Kawazoe, S. Nakashima, W. Nomura, M. Ohtsu, 
{\it Abstracts of the 22nd International Display Workshops}  (Dec. 9-11, 2015, Otsu, Japan, PRJ3-1).
%
\bibitem{DP01_Kawazoe}
T. Kawazoe, K. Kobayashi, S. Takubo, and M. Ohtsu, 
J. Chem. Phys.  {\bf 122,} 024715-1 
(2005).
\bibitem{DP02_Kobayashi}
K. Kobayashi, T. Kawazoe, and M. Ohtsu, 
IEEE Trans. Nanotech. {\bf 4,} 517
(2005).
%
\bibitem{OhtsuDP} M. Ohtsu, 
{\it Dressed Photons: Concepts of Light-Matter Fusion Technology}
 (Springer-Verlag, Berlin, Heidelberg, 2014).
%
 \bibitem{Cho02} K. Cho,  
 {\it Reconstruction of Macroscopic Maxwell Equations}
 (Springer-Verlag, Berlin, Heidelberg, 2010).
%
\bibitem{FDTD} 
K. Yee, 
{\it IEEE Transactions on Antennas and Propagation} {\bf14,}  302
(1966).
%
\bibitem{Keller01} O. Keller, 
{\it Quantum Theory of Near-Field Electrocynamics} (Springer-Verlag, Berlin, Heidelberg, 2011).
%
\bibitem{Cho01} K. Cho,   
{J. Phys. Condens. Matter} {\bf 20,}175202
(2008).
%
\bibitem{Cho03} K. Cho,    
{\it Optical Response of Nanostructures} (Springer-Verlag, Berlin, Heidelberg, 2003).
\end{thebibliography}
\end{document}